\documentclass[journal, letterpaper, 12pt, onecolumn]{IEEEtran}

\usepackage[dvips]{graphicx}
\usepackage{amsmath}
\usepackage{amssymb}
\usepackage{float}
\usepackage[nospace]{cite}
\usepackage{subfigure}
\usepackage{color}

\usepackage{tabularx}
\usepackage[dvips]{graphicx}
\usepackage{subfigure}
\usepackage[nospace]{cite}
\usepackage{comment}
\usepackage{mathrsfs}
\usepackage{array}
\usepackage{amsmath}
\usepackage{amssymb}
\usepackage{mdwmath}
\usepackage[english]{babel}
\usepackage{mathtools}
\usepackage{mdwtab}
\usepackage{amsthm}
\usepackage{textcomp}
\usepackage{times}
\usepackage{url}
\usepackage{bm}
\usepackage{algorithmic}
\usepackage[vlined,boxed,commentsnumbered,ruled]{algorithm2e}

\makeatletter
\renewcommand{\fnum@figure}{Fig.~\thefigure}
\makeatother

\IEEEoverridecommandlockouts
\graphicspath{{./}{./Figs/}}

\newtheorem{lemma}{\textbf{Lemma}}

\newtheorem{theorem}{\textbf{Theorem}}

\newcommand{\figtag}{\textrm{Fig.}}

\newcommand{\sectag}{\textrm{Section}}

\newcommand{\bms}{{\bm s}}
\newcommand{\bmbars}{ $\bar{\bms}$}
\newcommand{\eg}{{\it e.g.,~}}

\newcommand{\etc}{{\it etc.}}
\newcommand{\ie}{{\it i.e.,~}}

\newcommand{\algotag}{\textrm{Algorithm}}

\graphicspath{{./}{./figs/}}

\begin{document}

\title{\LARGE SOARAN: A Service-oriented Architecture for Radio Access Network Sharing in Evolving Mobile Networks\vspace{1em}}

\author{\small\IEEEauthorblockN{Jun He, and Wei Song, {\em Member, IEEE}}\\
\IEEEauthorblockA{Faculty of Computer Science\\
University of New Brunswick, Fredericton, Canada\\
Emails: \{jhe2, wsong\}@unb.ca}\\

\vspace{-4em}%
}

\maketitle

\begin{abstract}
Mobile networks are undergoing fast evolution to software-defined networking (SDN)
infrastructure in order to accommodate the ever-growing mobile traffic 
and overcome the network management nightmares caused by unremitting acceleration in technology innovations and evolution of the service market.
Enabled by virtualized network functionalities, evolving carrier wireless networks tend to share radio access network (RAN) 
among multiple (virtual) network operators so as to increase network capacity and reduce expenses.
However, existing RAN sharing models are operator-oriented, 
which expose extensive resource details,
\eg infrastructure and spectrum,
to participating network operators for resource-sharing purposes.
These old-fashioned models violate the design principles of SDN abstraction and are infeasible to manage the thriving traffic of on-demand customized services.
This paper presents SOARAN, a service-oriented framework for RAN sharing in mobile networks evolving from LTE/LTE advanced 
to software-defined carrier wireless networks(SD-CWNs),
which decouples network operators from radio resource by providing application-level differentiated services.
SOARAN defines a serial of abstract applications with distinct Quality of Experience (QoE) requirements.
The central controller periodically computes application-level resource allocation 
for each radio element with respect to runtime traffic demands and channel conditions,
and disseminate these allocation decisions as service-oriented policies to respect element.
The radio elements then independently determine flow-level resource allocation within each application to accomplish these  policies.
We formulate the application-level resource allocation as an optimization problem and 
develop a fast algorithm to solve it with a provably approximate guarantee.
The efficacy of SORAN is validated through theoretical analysis and computer simulations.
We also show that SORAN is in line with the design of SD-CWNs.
\vspace{-1em}%
\end{abstract}

\begin{IEEEkeywords}
Radio access network, RAN sharing, software-defined RAN, resource virtualization, network abstraction
\end{IEEEkeywords}

\bstctlcite{IEEEexample:BSTcontrol}%

\section{Introduction}\label{sec:introduction}
Nowadays, the mobile system has become a service-oriented platform 
swamped with millions of applications providing 
differential services to data-hungry devices whose owners increasingly regard
the ubiquitous quality-guaranteed network services as a human right, regardless of 
traffic overburdening of the network and high costs of system upgrading.
Sharing radio access networks (RANs) among mobile network operators (MNOs) and mobile virtual network operators (MVNOs)
is not only a promising way to expand system capacity and reduce both capital expenses (CAPEX) and operational expenses (OPEX)\cite{gsmareport},
but also an inevitable trend of carrier wireless networks facing the harsh realities of unremitting growth of traffic demand and continuously declining unit-data revenue.
MVNOs are a set of service providers (SPs) or network resellers, \eg content providers (CPs), 
who do not own network infrastructure but share MNOs' RANs in a rental manner based on service level agreements (SLAs) with them.
Along with the proliferation of smart devices, 
MVNOs play an essential role in enriching mobile networks with innovative applications, differentiated services,
and prompting subscribers' engagements\cite{telcordiawhite}, 
shifting mobile networks from operator-oriented systems to service-oriented systems.
However, despite the rapid development in software defined carrier wireless networks (SD-CWNs), 
service-oriented RAN sharing remains unexplored. 

\begin{figure}[t]
\centering
\includegraphics[width=0.65\linewidth]{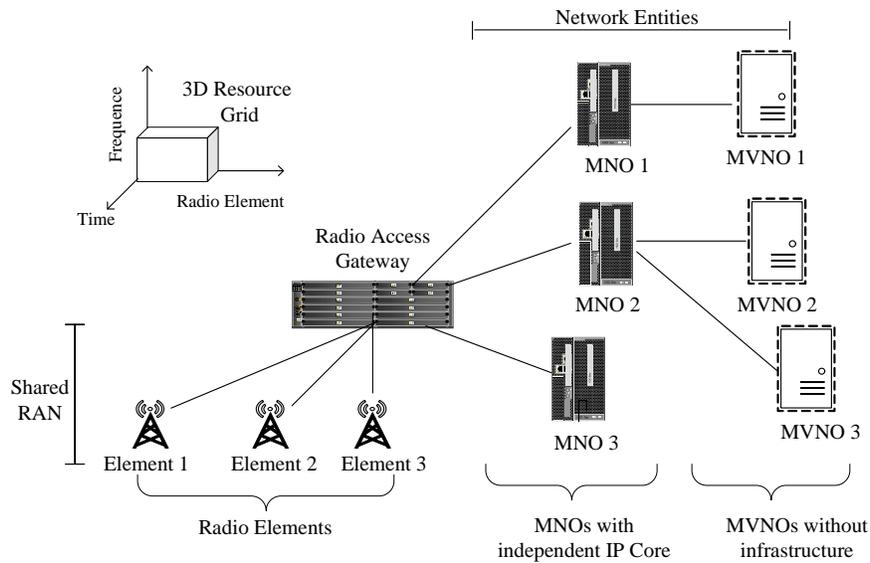}
\caption{RAN Sharing Network Model.}
\label{fig:lte}
\end{figure}

The oldest form of RAN sharing in MNOs is to accommodate ``foreign'' subscribers roaming from other networks. 
The Long-Term Evolution (LTE) system of the 3rd Generation Partnership Project (3GPP) supports RAN sharing among 
core network operators in either a gateway core network (GWCN) configuration or a multi-operator core network (MOCN) configuration 
with shared or independent mobility management entities (MMEs), respectively.
\figtag~\ref{fig:lte} shows the network model of sharing RAN we study in this paper.
MNOs share the RAN through a radio access gateway, \eg the serving gateway in LTE or the access service network (ASN) gateway in WiMAX,
and provide RAN access to MVNOs via their IP cores.
Radio elements refer to base-stations, \eg eNodeB, pico/micro cells,
that are managed by a centralized controller at the RAN gateway
and provide radio access to subscribers.

As mobile networks are merging into the cloud, RANs are also undergoing fast evolution to SD-CWNs with virtualized radio resource\cite{hoffmann2011network}.
With the promotion of scalability and manageability, virtualized RANs have developed maturing methods for resource slicing  and frame scheduling,
which eases the resource
management at the controller\cite{katti2014radiovisor,kokku2010nvs}.
As depicted in \figtag~\ref{fig:lte}, the radio resource over the RAN
is abstracted in a configurable {\em 3-dimensional resource grid} of radio element index, frequency and time\cite{katti2014radiovisor}.
The central controller has a view of one virtual ``big'' base-station upon which radio resource
is slicable and allocatable via the northbound application programming interfaces (APIs) that the virtualized RAN provides to the controller.

State-of-the-art resource schedulers, \eg \cite{mahindranetwork,kokku2010nvs}, 
divide resource among operators sharing the RAN based on SLAs with operator isolation in a resource-reservation manner. 
That is, SLAs specify the resource shares of each operator either on a per-base-station basis (\eg \cite{kokku2010nvs}) or on a RAN basis (\eg \cite{mahindranetwork}).
For instance, in a network with 2 operators, operator 1 reserves $30\%$ of the resource (overall or per-base-station) while operator 2 takes $70\%$.
These share ratios could also be a range, \eg minimal $20\%$ and maximal $35\%$, to enable adaptive resource scheduling according to data traffic\cite{mahindranetwork}.
The allocation decisions  made at the controller are then applied by lower-layer frame schedulers.
However, due to the following concerns, we argue that these operator-oriented designs 
are against the principles of network virtualization and will soon become infeasible in the expanding mobile networks.
\begin{itemize}
\item These methods expose extensive details of RAN to the operators, \eg the number, distribution and capacity of radio elements.
Even though the resource share is given in percentages,
for pricing and budget purposes, one operator needs to know the coverage of the RAN and the bandwidth it provides.
Therefore, these operator-oriented approaches will make the network management and RAN upgrade even more complicated as the RAN or the number of operators grows.
\item To fulfill SLAs, the controller requires the ownership information of each data flow, 
which can be retrieved in no way but by conducting deep packet inspections (DPIs).
Apparently, the overhead of DPI per flow would be intolerable.
\item It is hard to manage Quality of Experience (QoE) in these models. 
Each operator provides a set of services (regarded as applications hereafter) with differentiated QoE requirements.
Since the resource is allocated at the operator level, each operator independently needs to manage QoE of its own applications, 
\eg using bearers \cite{sesia2009lte},
which not only plunges QoE support at the RAN gateway into chaos, but also produces an aggregate resource utility arbitrarily suboptimal.
\item These existing works determine detailed resource allocation (\eg per flow) at the central controller, which is incompatible to new developments
in RAN with heterogeneous radio elements due to the overwhelming overhead of reporting wireless channel details.
For example, only symmetric base-stations are considered in \cite{mahindranetwork}.
\end{itemize}

To address these problems, we proposed an application sepecified  RAN sharing architecture in \cite{jun2015icc}.
In this paper,
we extend the intention of \cite{jun2015icc} and re-define the RAN sharing model, namely SOARAN, a service-oriented architecture for RAN sharing in mobile networks.
Instead of promising certain amont of resource for each operator, 
SOARAN defines a configurable set of abstract applications with respect to differentiated QoE requirements such that 
operators can map their concrete applications to the abstract ones and
determine application-specified bandwidth they need.
Operators can now focus on their application-level demands through an application-abstraction layer provided by SOARAN,
while SOARAN takes care of lower-layer resource allocation.
In this way, 
SOARAN decouples operators from radio resource allocation,
keeps the upgrade of RAN facilities and resource allocation transparent to operators
and enables better resource virtualization.

A new model of SLAs is defined based on the promised services for each operator.
The charging policy is then determined by the service package each operator purchases,
\eg in the form of a series of (application, bandwidth) tuples.
Such service-oriented SLA model makes a step towards merging RANs into the cloud.
SOARAN also develops an optimization framework and a fast algorithm to optmize the resource allocation.
The optimization framework takes average resource-to-bandwidth conversion ratios reported by radio elements
as input and computes the optimal resource allocation among applications for each radio element.
With negotiable overhead, the central controller determines the optimal resource allocation on the application level.

The rest of paper is organized as follows.
Section II describes the design of SOARAN.
The kernel resource allocation algorithm is presented in Section III.
Section IV provides the numerical results and Section V concludes the paper.

\section{Design of SOARAN}\label{sec:overview}

\begin{figure}[t]
\centering
\includegraphics[width=0.40\linewidth]{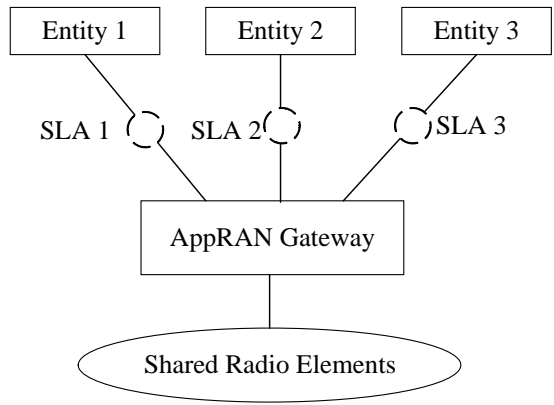}
\caption{SOARAN Network Model.}
\label{fig:appnet}
\end{figure}

SOARAN flattens the RAN sharing structure and re-models it in \figtag~\ref{fig:appnet}.
Network operators including MNOs, MVNOs, are regarded as {\em entities} driving data flows to the
SOARAN gateway, where the flows are differentially treated according to SLA configurations.
An MNO with several virtual operators attached to it (see \figtag~\ref{fig:lte}) can be treated as one entity or several entities as it describes,
while the details inside are kept transparent to SOARAN, granting more flexibility to the MNO.

\subsection{The SOARAN Model}
\begin{figure}[t]
\centering
\includegraphics[width=0.65\linewidth]{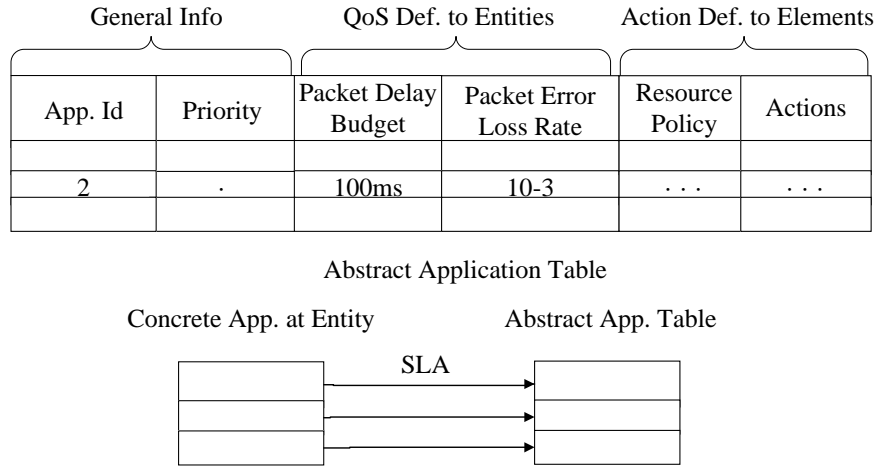}
\caption{Definition of Abstract Application and SLA Model.}
\label{fig:appsla}
\end{figure}

SOARAN defines a series of {\em abstract applications} with respect to
differentiated QoE levels, which can be readily supported using RAN ``bearers'' in 3GPP systems\cite{sesia2009lte}.
\figtag~\ref{fig:appsla} shows an example of the abstract application table.
The description of an abstract application consists of identification information (id and priority), QoE guarantees to entities (delay, packet loss rate, \etc),
action information to radio elements (resource policy, network actions, \etc), and (possibly) unit pricing information.
An SLA then indicates how to map concrete applications to abstract applications and the bandwidth demand upon each abstract application.
SOARAN thus adapts differentiated services\cite{Nichols1998} supported at respective entities
to the abstract application set.
In this way, SOARAN is able to react quickly to fast-growing emerging applications, which will become the new norm in future networks\cite{horizon2020},
by adding entries at the abstract application table. 
SOARAN also eases network management and resource allocation by abstracting numerous external nonuniform services in a controllable set.
As a result, entities are only required to determine the types and the bandwidths of abstract applications they need
on a more trackable and readable pricing system produced by the re-modeled SLA.

According to the bandwidth requirements of respective abstract applications gathered from SLAs,
the SOARAN controller configures a lower bound
and an upper bound of resource available to each application to enforce application-level isolation,
while the resource within the bounds is adjustable and periodically allocated to each application at the time order of several seconds
to promote resource utility with respect to dynamic traffic demands and varing wireless channel conditions.

\subsection{The SOARAN Software Architecture}
\begin{figure}[t]
\centering
\includegraphics[width=0.65\linewidth]{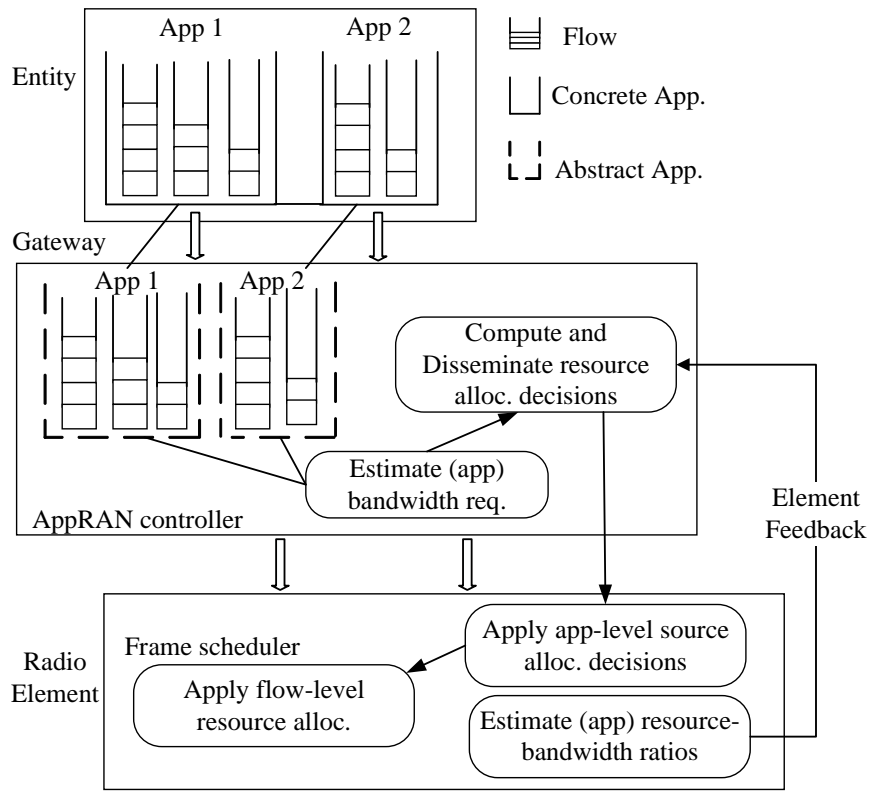}
\caption{Software Architecture of SOARAN.}
\label{fig:appsoft}
\end{figure}
The software architecture of SOARAN is illustrated in \figtag~\ref{fig:appsoft}.
Flows of concrete applications in entities are mapped to abstract applications upon entering
the RAN gateway according to respective SLAs.
Inside the SOARAN gateway, per time period $\tau$ (in the order of seconds),
the controller estimates the bandwidth requirement of each application (possibly) through history analysis.
On the same periodic scale, each radio element estimates the average resource-bandwidth ratio associated with each application,
\ie average resource per unit data rate, capturing the average channel condition of corresponding application,
 and sends these ratios to the controller as element feedback.
The rationale behind this is that the resource to support unit data rate is jointly determined by QoE requirements (indicated by types of applications)
and wireless channel conditions.

Together with the estimated bandwidth requirements, the controller computes the dynamic resource requirement of each application and determines
how the resource should be divided among these applications. 
The calculated resource allocation decisions are then disseminated to respective elements as resource {\em policies}.
Rather than specifying resource allocation for each flow,
SOARAN creates resource policies at the application level while granting the flexibility to elements on how to accomplish these policies.
We comment that this design is
in line with the principle of the SDN architecture\cite{nunes2014survey}.
As the radio element has more accurate information of channel conditions as well as fine-grained power
management and maturing modulation and coding technologies\cite{han2008resource},
allowing elements to optimize local resource allocation will further improve resource utility and enable heterogeneous network deployment\cite{han2008resource}.

\section{Optimal Resource Allocation in SOARAN} \label{sec:resource}
The resource allocation scheduler is the kernel of SOARAN, running at the logically central controller (see \figtag~ \ref{fig:appsoft}), which
periodically computes the resource distribution among the applications with respect to current network conditions and system configurations.

\subsection{The Optimization Framework}

\subsubsection{Formulation}
We target a RAN configuration with a set of $I$ radio elements supporting a set of $K$ applications.
Each radio element $i\in I$ has a resource amount $B_i$, representing the available radio resource at the element,
abstracted from the three-dimensional resource grid.
The aggregate radio resource over all elements is denoted by $B$,
\ie $B=\sum_{i\in I} B_i$.
For isolation purpose, each application $k\in K$ reserves a minimum resource of $L^k$ irrespective to traffic demands,
while it can use up to $M^k$ resource to improve its performance, where $L^k\le M^k \le B$.
Similarly, the system also configures a lower bound $l_i^k$ and a upper bound $m_i^k$ for
resource allocation to application $k$ at element $i$ to enable element-level isolation,
accordingly, $L^k=\sum_{i\in I} l_i^k$ and $M^k=\sum_{i\in I} m_i^k$.

Let $s_i^k$ be the amount of resource allocated to application $k$ at element $i$, $i \in I,  k\in K$.
Now for each time period $\tau$, we aim to maximize the overall resource allocation gain or utility while
assuring that the resource used by each application is bounded according to preset configurations.
Defining the utility function $u_i^k(\cdotp)$, we formulate the following optimization problem:

\begin{equation}
\begin{split}
\label{eq:opt}
\displaystyle\max &\sum_{k \in K} \sum_{i\in I} u_i^k(s_i^k) \\
\textrm{s. t.}~  &\sum_{k\in K} s_i^k\le B_i, ~ \forall i\in I \\
&\sum_{i\in I} s_i^k \le M^k, ~ \forall k \in K \\
&\sum_{i\in I} s_i^k \ge L^k,~ \forall k\in K\\
\textrm{var.}~& 0 \le l_i^k \le s_i^k \le m_i^k, ~ \forall i\in I, k \in K.
\end{split}
\end{equation}
The first constraint indicates the resource limit at each radio element $i$.
The second and third constraints impose the upper bound and lower bound of resource that can be allocated to each
application $k$ over the RAN, respectively.
The per-element resource upper bound and lower bound for each application are imposed by the last constraint.

A similar optimization problem is formulated in \cite{mahindranetwork}, which attempts to maximize
the aggregate utility of allocating network-wide radio resource proportionally to entities or mobile virtual network operators.
However, they only consider the symmetric scenario by assuming that all base-stations (termed as radio elements in this paper) possess the same amount of resource,
which prevents their framework from scaling to the complicated networking reality nowadays with heterogeneous wireless elements.
On the contrary, our formulation directly  addresses the amount of radio resource, allowing heterogeneity of elements.
Providing the information of available resource at each element, these resource amounts are readily to be converted to
percentages for implementation purposes.
Therefore, problem (\ref{eq:opt}) fundamentally differs from the model formulated in \cite{mahindranetwork}.

\subsubsection{Utility Function and Demand Estimation}
The utility function $u_i^k(\cdot)$ can be a linear function or any concave function
following the law of diminishing marginal utility,
representing the utility value of allocated resource to application $k$ at element $i$.
The following equations show examples of a linear function and
a logarithmic function drawn from the proportional fairness principles
defined in \cite{tychogiorgos2012utility}:
\begin{align}
\nonumber
  u_i^k(s_i^k)&= w_i^k\cdot d_i^k \cdot s_i^k, ~~~~~~~\mathrm{or}\\
\label{eq:log}
 u_i^k(s_i^k)&= w_i^k\cdot d_i^k \cdot \log (s_i^k),
\end{align}
where $w_i^k$ is the utility weight, $d_i^k$ is the resource demand for application $k$ at element $i$ in current period.

In conventional RAN sharing models, it is intractable to estimate resource demand $d_i^k$ with respect to
distinct entity, let alone to support differentiated QoE for different applications within one entity.
For one thing, the network-level resource allocator has no information of the ownership of flows,
\ie to which entity each flow belongs.
This forces the system either to use the off-line, long-term estimation of ``average'' demands, or to employ
deep packet inspection (DPI) to extract application-level information from flows.
The former lacks accuracy, while the latter apparently introduces an intolerable computational overhead.
For another, translating flow-level bandwidth demands to radio resource demands requires the information of modulation coding schemes (MCSs)
selected for each flow transmission, which in turn relys on element-level details of users' channel conditions and MCS adaptation schemes\cite{han2008resource}.

Bandwidth demands of ongoing flows in SOARAN are irrespective of the entities they belong to, requiring no DPI operations.
Mobile systems usually support Differentiated Services (DiffServ)\cite{Nichols1998} in their IP backbones for QoE management, \eg evolved packet system (EPS) bearers in
LTE systems \cite{sesia2009lte}. With DiffServ, SOARAN easily determines to which application a flow belongs by checking the QoE class identifier (QCI) attached to the flow.
Moreover, together with the knowledge of channel conditions and scheduling algorithms,
each radio element is ready to translate bandwidth demands of any application to resource demands.
We define the average bandwidth-resource translating ratio for application $k$ at radio element $i$ as
$p_i^k = \frac{resource~to ~ support~ application ~k~with~ QoE  }{bandwidth~demand~for~application~k}$,
which is reported to the central controller for resource-demand estimation.
Here, we note that such translating ratio might not reveal the ``real'' relation between bandwidth and resource demands in the cases with significant flow fluctuation,
\eg the traffic demand of the user with the worst channel condition soars for the next time period $\tau$,
or the channel condition of a heavy-traffic user significantly changes.
Yet we argue that our approach remains effective.
This is because: (1) For a short time period of $\tau$, it is less likely to have large fluctuation. Even with large fluctuation,
the system only experiences suboptimal resource allocation for at most $\tau$ time; and (2) In SOARAN, we compute the resources allocated at each element to distinct applications. Therefore, the fluctuation can be mitigated or shaped by employing adaptive MCS schemes\cite{chung2001degrees} at radio elements. That is, given an application and allocated resource, the element runs
a second-phase resource allocation to distribute resource among flows with accurate channel state information.

\subsubsection{Problem Hardness}
The utility function is either a linear function or a concave function as shown in  (\ref{eq:log}), resulting 
in a linear programming (LP) model and a nonlinear programming (NLP) model for problem (\ref{eq:opt}) respectively.
However, in both models, problem (\ref{eq:opt}) has a prohibitively large size for a direct solution from state-of-the-art LP/NLP solvers, \eg CPLEX, OPT++.
To have a rough understanding, a production mobile system usually has $O(10^5)$ radio elements and
supports $O(10^2)$ applications. Therefore, the rudimentary size of problem (\ref{eq:opt})
is with $O(10^7)$ variables and $O(10^7)$ constraints (see the last constraint of problem (\ref{eq:opt})).
Therefore, a fast algorithm with approximate guarantees is more desirable.

\subsection{The Approximate Algorithm}

\subsubsection{Main Procedure}
We employ the Barrier method from \cite{boyd2009convex} and solve problem (\ref{eq:opt}) via an interior-point approach.
For ease of presentation, let ${\bm s}$ be the vector of variables $\{s_i^k| i\in I, k\in K\}$.
We define the logarithmic barrier function as
\begin{equation}
\label{eq:barrier}
\begin{split}
 \phi ({\bm s})= &\sum_{i\in I} \log(B_i - \sum_{k\in K} s_i^k) + \sum_{k\in K} \log(M^k - \sum_{i\in I} s_i^k) \\
   &+ \sum_{k\in K} \log(\sum_{i\in I} s_i^k -L^k).
\end{split}
\end{equation}

We denote the objective of problem (\ref{eq:opt}) by $u(\bms)=\sum_{k \in K} \sum_{i\in I} u_i^k(s_i^k)$.
We then introduce a multiplier $t$ and consider the following problem:
\begin{equation}
\label{eq:interior}
\begin{split}
\max~ & t\cdot u(\bms) + \phi(\bms) \\
\textrm{s. t.}~ & l_i^k \le s_i^k \le m_i^k, ~ \forall i\in I, k \in K.
\end{split}
\end{equation}

The main procedure of our algorithm is listed in \algotag~\ref{alg:barrier}.
The procedure follows a typical route of the Barrier method, while we develop a tighter bound.
Starting from a feasible point, it iteratively solves a sequence of problem (\ref{eq:interior})
with increasing $t$ till $t\ge (B+|K|)/\epsilon$ (to be discussed later).
For simplicity, we set ${\bm s}_0 = {\bm l}$ to be the initial starting point, providing that ${\bm l}$ is a feasible solution to problem (\ref{eq:opt})
under proper configurations, \ie $\sum_k l_i^k\le B_i,\forall i\in I$.
Line \ref{line:newton} therein is called an inner loop for solving the optimization problem  (\ref{eq:interior}) with bounded variables.
We refer readers to \cite{boyd2009convex} for the details of the inner loop, \eg the Newton's method.
Each solution found in line \ref{line:newton} is then used as a new starting point for the next iteration in the outer loop.
Here, $\mu$ is a parameter involving a trade-off in the number of iterations of the inner and outer loops. Details on selecting $\mu$ can also be found in \cite{boyd2009convex}.

\begin{algorithm}
\caption {Barrier method for problem (\ref{eq:opt}).}
\label{alg:barrier}
\begin{algorithmic}[1]
\STATE  Start with an interior feasible point ${{\bm s}_0}$;
\WHILE{$(B+|K|)/t>\epsilon$ $(\epsilon> 0)$}
\STATE ${{\bm s}}:={{\bm s}}_0,~t:=t_0$, where $t_0>0$;
\STATE With starting point ${\bm x}$, solve (\ref{eq:interior})
via a gradient-based method and output the solution ${\bm x}^*$. \label{line:newton}
\STATE ${\bm s}:={\bm s}^*,~ t:=t \cdot \mu$, where $\mu>1$;
\ENDWHILE
\end{algorithmic}
\end{algorithm}

\subsubsection{The Approximate Result} \label{sssec:app}
We now develop the theoretical basis for \algotag~\ref{alg:barrier}.
Applying the duality analysis, we have the following conclusions.
\begin{lemma}\label{lem:approximate}
If problem (\ref{eq:interior}) can be optimally solved,
then we can find a solution to problem (\ref{eq:opt}) that is at most $\epsilon-$suboptimal, for any $\epsilon >0$.
In other words, let $u^*$ be the optimal value of problem (\ref{eq:opt}) and \bmbars ~ be the optimal solution to problem (\ref{eq:interior}).
By setting $t\ge (B+|K|)/\epsilon$, we have
$$u^*\le u(\bar{\bms}) +\epsilon.$$
\end{lemma}
\begin{theorem}
\label{th:app}
 For any $\epsilon >0$, \algotag~\ref{alg:barrier} obtains a solution to problem (\ref{eq:opt}), which is at most $\epsilon-$suboptimal.
\end{theorem}

\section{Simulation and Numerical Results}\label{sec:simulation}
In this section, we study the SOARAN model through extensive computer simulations.
We focus on the application-level resource allocation at the RAN gateway, while
we assume that certain flow-level resource allocation schemes, \eg \cite{han2008resource},
are adopted by each radio element and radio resource therein is abstracted and
represented as a non-negative real value using the technologies in \cite{gudipati2013softran, katti2014radiovisor}.

\subsection{Simulation Setup} \label{ssec:setup}
We simulate a RAN system with $1,000$ radio elements shared by $20$ entities.
SOARAN defines $100$ abstract applications with resource requirement factors uniformly selected within $[0.1,2]$ per unit data rate (Mbps),
representing respective QoE guarantees.
We assume that all these applications are supported over all entities and radio elements to exclude the complexity of SLAs from our simulations.
Given that the information of channel conditions is available to elements,
the average resource-bandwidth multipliers are uniformly generated from $[1.0,2.0]$, 
resulting in a resource-bandwidth ratio range of $[0.1,4.0]$ (jointly determined by channel conditions and QoE requirements).
The available resource at elements is abstracted as real values randomly generated from $[100.0,300.0]$ 
with the mean value of $200.0$ over all elements. 
This setting is to ensure that the logarithmic utility function results in a non-negative value in all cases, representing
proportional resource capacities that can be arbitrarily scaled up/down over the system.
Two most data-consuming applications, $m$ and $n$, \eg video streaming and FTP file downloading, 
can use at least $5\%$ and up to $40\%$ of the aggregate resource, while other applications equally share the rest resource.

For comparison, we align SOARAN with two alternative operator-oriented resource allocation schemes, 
termed as {\em Per-Base-station Reservation} (Per-Bs-Rsv)\cite{kokku2010nvs} and
{\em Network Reservation} (Net-Rsv)\cite{mahindranetwork}, in which the utility is calculated over flows instead of entities as in the literature.
In Net-Rsv, each entity reserves $2\%$ of the aggregate resource and can use up to $10\%$, 
while in PerBs-Rsv, each entity reserves $5\%$ resource at each radio element.
For comparison fairness, the utility is calculated over flows for all three schemes to be comparable, 
since the accumulated utility either over applications or entities can be decomposed in the form of flows.
The basic system load contains $5,000$ flows randomly generated from all applications across all radio elements, 
the bandwidth demands of which are selected from
$0.1\sim 1$ Mbps to ensure that a feasible resource assignment can be reached by all schemes.
This load is then gradually increased by a controlled multiplier for different scenarios.
\begin{figure*}
\centering
\subfigure[Utility with Linear Function.]{\label{fig:linearu}\includegraphics[width=0.30\textwidth]{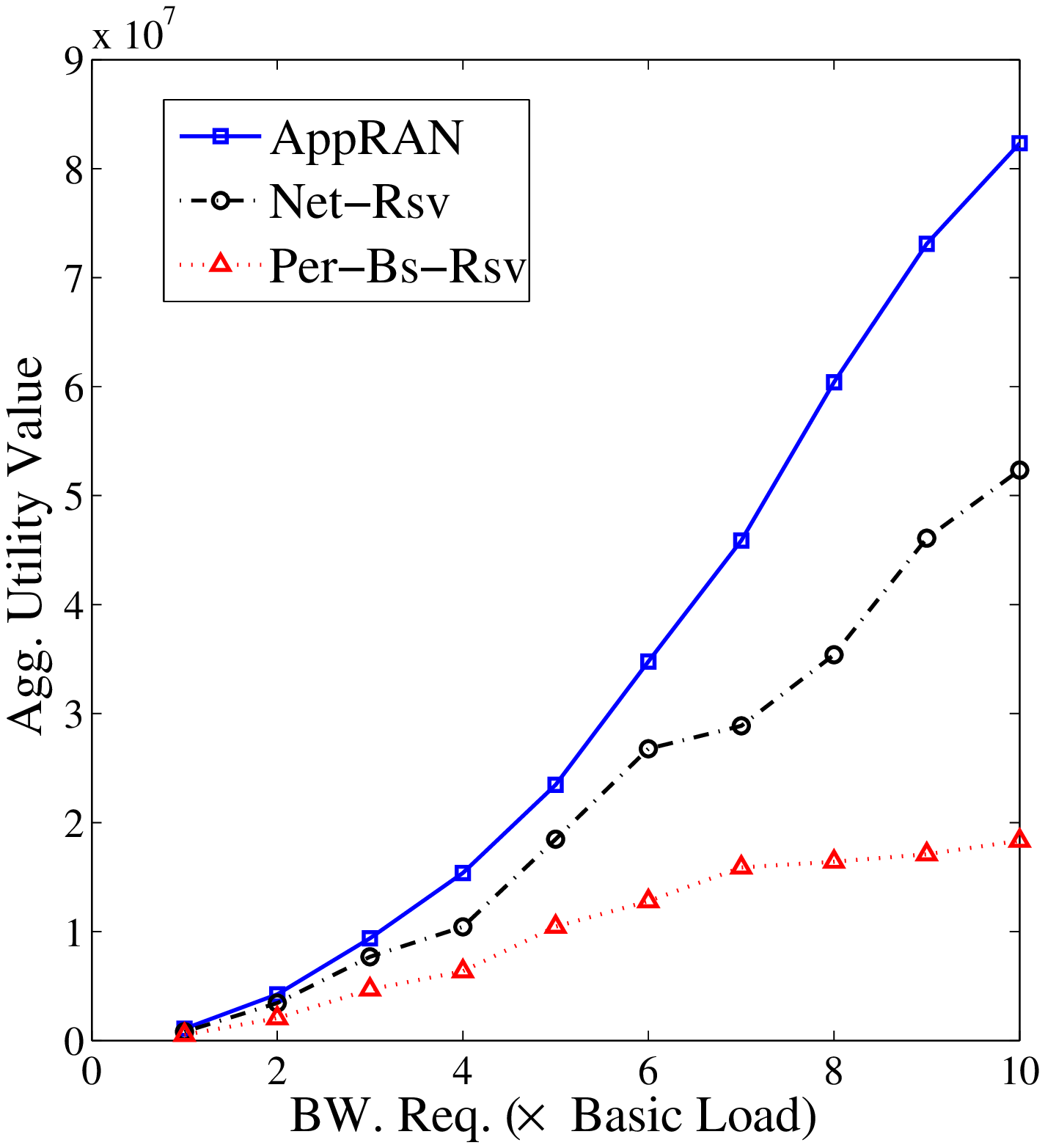}}
\subfigure[Utility with Log. Function.]{\label{fig:logu}\includegraphics[width=0.30\textwidth]{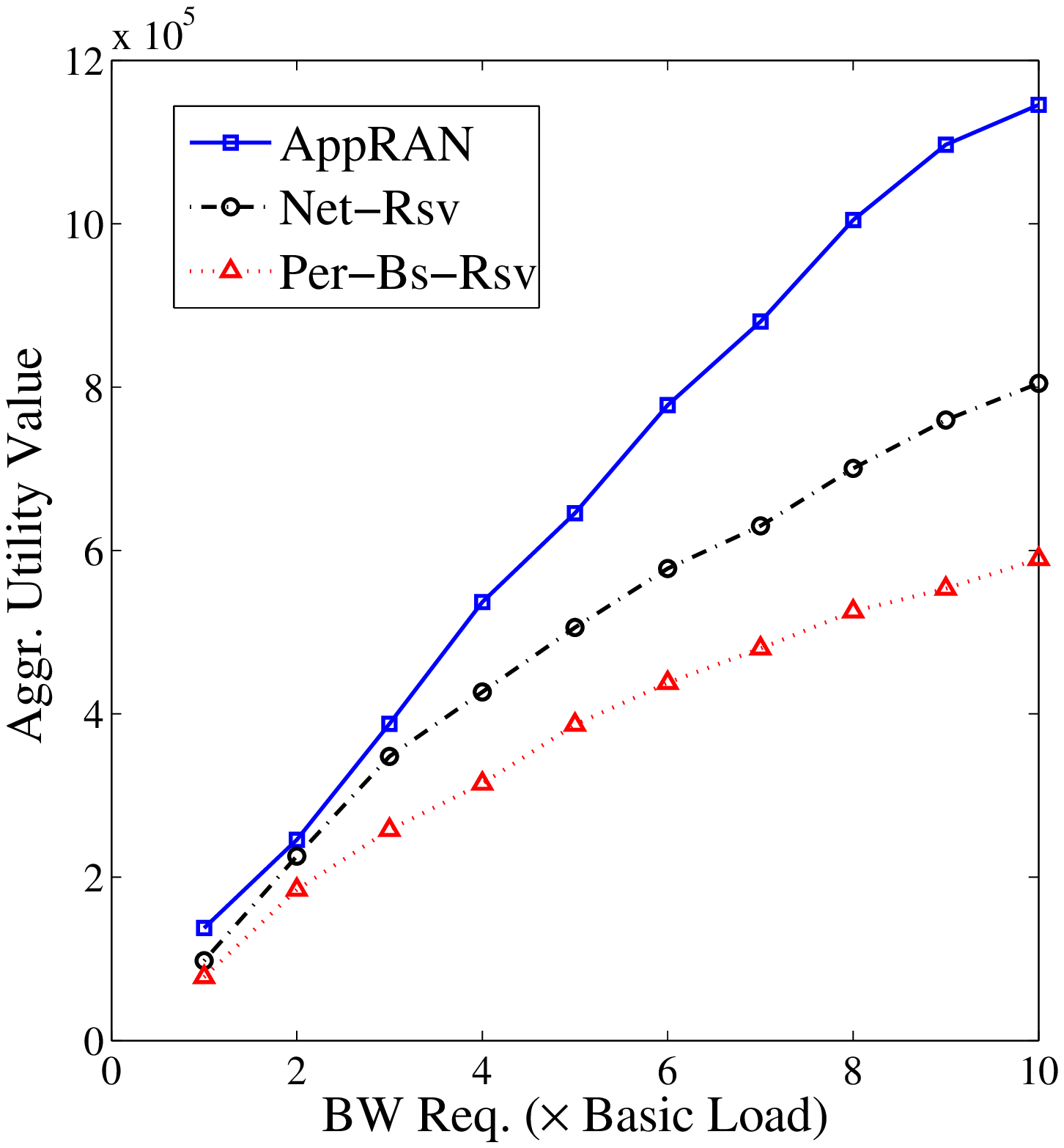}}
\subfigure[Resource Usage for Application $m$.]{\label{fig:resource}\includegraphics[width=0.30\textwidth]{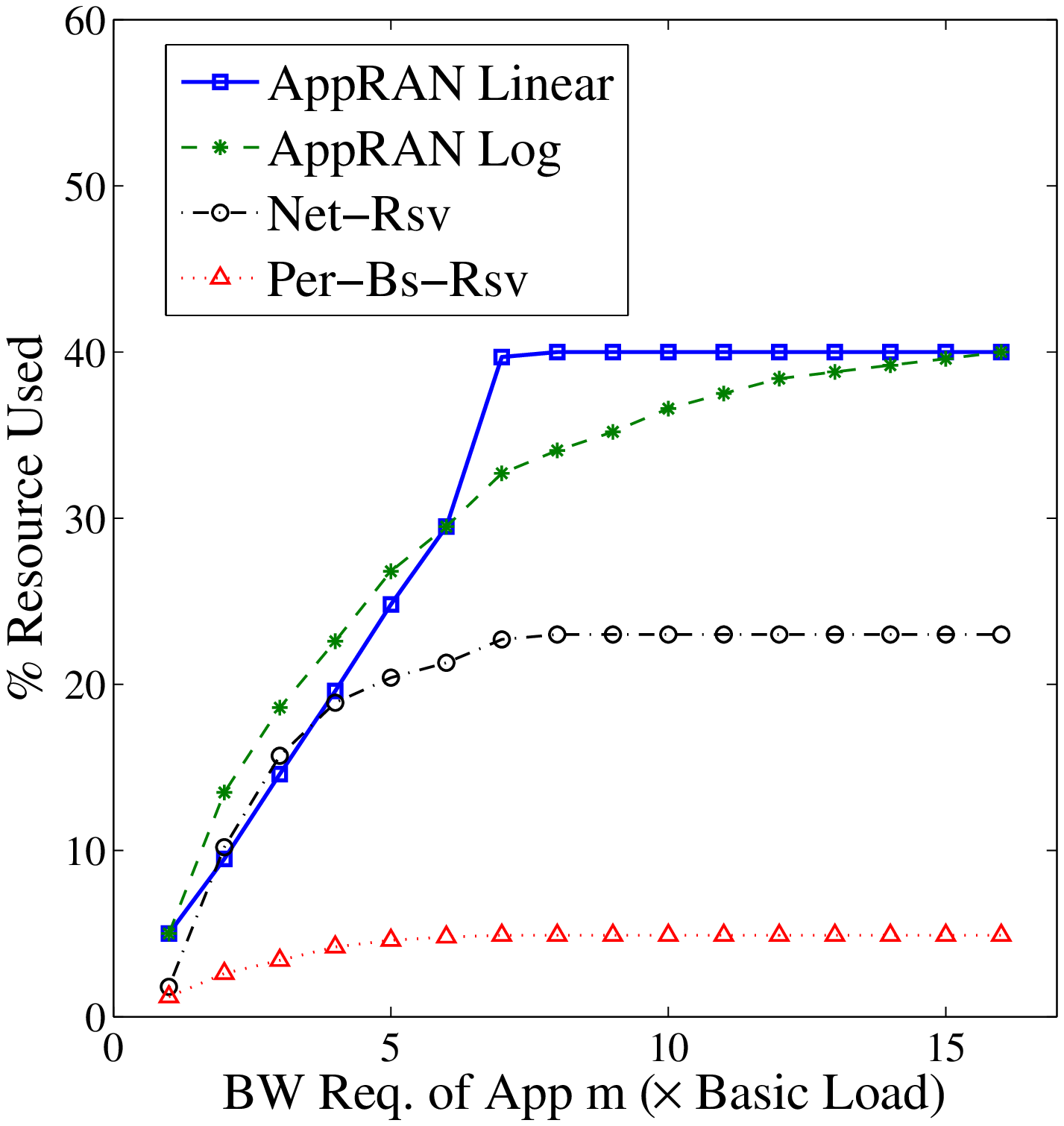}}
\caption{Performance comparison of SOARAN and operator-oriented alternatives. }
\label{fig:te:app}
\end{figure*}

\subsection{Utility Results}
For simplicity, we set unit utility weights $w_i^k=1$ $\forall i\in I$, $k\in K$.
We use the linear function and the logarithmic function (\ref{eq:log}) for
utility calculations and show results in \figtag~\ref{fig:linearu} and \figtag~\ref{fig:logu}, respectively.
In both scenarios, the system load increases step-by-step from $1$ to $10$ times of of the basic load.
With a linear utility function, the quadratic-like utility-growing curve of SOARAN in \figtag~\ref{fig:linearu} shows that
SOARAN tends to allocate resource linearly to corresponding bandwidth demands when the system is under a low to moderate load.
This growth is flattened with a logarithmic utility function (\figtag~\ref{fig:logu}), which also considers the fairness among applications.
In both cases, SOARAN obtains a utility objective which significantly outperforms the Net-Rsv scheme (up to $40\%$) and the Per-Bs-Rsv scheme (up to $200\%$).
This confirms the conclusion that resource reservation over entities immensely limits the RAN sharing performance with multiple QoE-differentiated applications.
\begin{figure}[t]
\centering
\includegraphics[width=0.45\linewidth, height=3.0in]{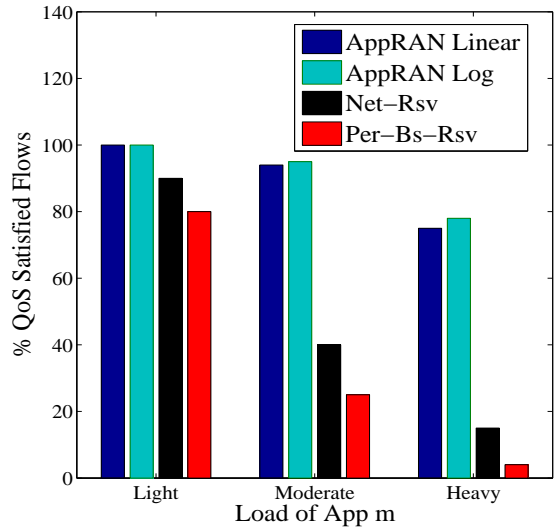}
\caption{Comparsions of QoE Satisfied Flows.}
\label{fig:qos}
\end{figure}

\subsection{QoE Results}
In this simulation, we study the QoE performance of different schemes
by gradually increasing the bandwidth demands of application $m$, one of the most data-consuming applications.
Beyond the basic load setup in \sectag~\ref{ssec:setup}, we add $2,000$ extra flows of application $m$ from $5$ entities to randomly selected $200$ radio elements
 with a mean basic bandwidth demand of $1.0$ Mbps.
The load of application $m$ is then iteratively increased from $1$ to  $15$ times of the basic load, while the demands of rest flows keep unchanged.
\figtag~\ref{fig:resource} shows the resource consumption of application $m$.
It indicates that SOARAN effectively adapts resource allocation for data-consuming applications as the traffic demands increase. In contrast,
as constrained by per-entity resource limits, both Net-Rsv and Per-Bs-Rsv result in significant resource under-utilization irrespective of idle resource in the system.
In SOARAN, we also observe that with the linear utility function, the resource usage grows more aggressively to its resource upper bound.

In \figtag~\ref{fig:qos}, we show the number of QoE satisfied flows in each case. Here, light, moderate 
and heavy loads correspond to $1$, $10$ and $15$ times of the basic load of application $m$.
We can see from \figtag~\ref{fig:qos} that SOARAN achieves similar QoE performance with the linear or logarithmic utility function.
However, with entity-oriented resource reservation, Net-Rsv cannot even fully support all flows 
with the least load and the performance deteriorates further as the load increases. 
Likewise, Per-Bs-Rsv produces the worst performance due to its strict entity-oriented resource constraints.

\section{Conclusion} \label{sec:conclusion}

In this paper, we propose SOARAN, an service-oriented framework for sharing RAN resource in evovling mobile networks.
SOARAN encapsulates the complexity of underlying resource management by
defining an application abstraction layer for entities and provides QoE-guaranteed services to them.
SOARAN centrally optimizes resource distribution among applications at each element, while
the decisions on allocating resource to flows are determined distributively at
each element with real-time channel conditions. 
By decoupling RAN sharing participants from radio resource, 
SOARAN is in line with the principles of SDN and enables better network abstraction.
A fast algorithm has been proposed and studied for service-level resource allocation. The simulation results demonstrate
significant performance improvement of SOARAN over entity-oriented schemes in terms of aggregate utility and QoE satisfaction.


\bibliographystyle{IEEEtran}
\bibliography{sharing}

\end{document}